\documentclass[fleqn,twoside]{article}
\usepackage{espcrc2}
\usepackage[polish]{babel}
\usepackage[OT4]{fontenc}
\usepackage[cp1250]{inputenc} 

\usepackage{graphicx}
\usepackage[figuresright]{rotating}

\newcommand{\AmS}{{\protect\the\textfont2
  A\kern-.1667em\lower.5ex\hbox{M}\kern-.125emS}}

\title{Tau polarization in charge current neutrino-nucleon Deep Inelastic Scattering}

\author{Krzysztof M. Graczyk\address[IFT]{Institute of Theoretical Physics \\
        University of Wroc\l aw\\
        pl. M. Borna 9, 50-204, Wroc\l aw, Poland}%
        \thanks{The autor is
supported by the Maxa Borna scholarship. This work was also
supported  by the KBN grant of number 105/E-344/SPB/ICARUS/P-03/DZ
211/2003-2005. } }
\begin{document}

\begin{abstract}
Numerical results for the degree of polarization of $\tau^-$
produced in (CC) neutrino - nucleon Deep Inelastic Scattering
(DIS) are presented. Calculations are done in the threshold
region, where the $\tau^-$ scattered by the small angles  can be
partially polarized. The cross sections and polarization are
calculated by using the GRV98 parton distributions functions
(PDF's) and the GRV98 with modifications of A. Bodek
at.al.~\cite{Bodek}.
  \vspace{1pc}
\end{abstract}

\maketitle

\section{Introduction}
The oscillation of muon to tau neutrinos is the most acceptable
solution of the atmospheric muon neutrino deficit measured in the
Super Kamionkande. Experiments such as ICARUS, OPERA  will verify
it by observation of tau  neutrinos  resulting from oscillations
of the CNGS muon neutrinos. The expected number of $\nu_\tau$
events will not be large.  Therefore a detailed analysis of
experimental data  based on a good theoretical description of the
neutrino-matter interaction will be necessary.

The tau leptons produced in the neutrino-matter interaction can
not be measured directly because of their short lifetime,
therefore their decay products must be detected. The large mass of
the tau lepton implies that it may not be fully
polarized~\cite{Smith,Albright}. The degree of its polarization is
one of the parameters of its  decay~\cite{Leader}. Hence it is
important to take into consideration the polarization properties
of $\tau$. It can play an important role in the analysis of
experimental data.

The polarization of the tau lepton produced in the neutrino -
matter interaction has been the subject of several
papers~\cite{japonczycy,rosjanie,Soffer,Graczyk}. In
\cite{japonczycy,rosjanie} the calculation were performed for
quasielastic neutrino-nucleon scattering, for single pion
production, and for deep inelastic scattering. The polarization
vector was obtained from the spin density matrix.

In the CNGS experiments the majority of neutrinos is detected by
the observations of the products of their inelastic scattering on
nucleons. Usually to describe the events with small values of
hadronic invariant mass some kind of the resonance model like
Rein-Sighal is used. Other events are described within the Deep
Inelastic Scattering formalism.

The tau leptons produced in the inelastic scattering are usually
characterized by higher degree of polarization~\cite{japonczycy}
but for small neutrino energies and small scattering angles the
taus may be partially polarized. In the present paper we focus on
this energy and scattering angle values.

The DIS formalism is described by the structure functions
expressed in terms of parton distribution functions  (PDF's). We
adopt the GRV98 PDF's from ~\cite{GRV98}, but  it seems that in
the region of small $Q^2$ they need to be modified. A. Bodek et.
al.~\cite{Bodek} proposed a modification based on the new
experimental data from JLab. We discuss how these modifications
influence the produced leptons degree of polarization and its
energy distribution.

\section{Theoretical description}
We consider the following process: \begin{footnotesize}
$$
\nu(k) + N(p)  \rightarrow  \tau^-(k',s^\mu) + X
$$
\end{footnotesize}
The produced tau lepton is polarized and is characterized by a
spin four-vector $s_\mu$ which satisfies in any frame the
relations: $ {k'}_\mu s^\mu =0$, $s_\mu^2 = -1.$ The cross section
is proportional to the contraction of the lepton tensor with the
hadron tensor: \begin{footnotesize}$$ d\sigma(k,q,s) \sim
L_{\mu\nu} W^{\mu\nu}.
$$\end{footnotesize}
We use the well known form of the hadron tensor
\hbox{\cite{Leader2}} \begin{footnotesize}
\begin{eqnarray}
W_{\mu\nu}& = & - g_{\mu\nu}F_1 + \frac{p_\tau p_\nu}{p\cdot q}F_2
-\mathrm{i}\frac{\epsilon_{\mu\nu\alpha\beta}p^\alpha
q^\beta}{2p\cdot q}F_3 \nonumber\\
\label{ten_hadronowy_LS} & &+\frac{q_\mu q_\nu}{p\cdot q}F_4 +
\frac{p_\mu q_\nu + p_\nu q_\mu}{2 p\cdot q}F _5
\end{eqnarray}\end{footnotesize}
$F_i$ (i=1,2,...,5) are the structure functions. The lepton
tensor:
 \begin{footnotesize}\begin{eqnarray*} L_{\mu\nu} & = & 8
\left(k_\mu {k'}_\nu +k_\nu {k'}_\mu -g_{\mu\nu}{k'}_\alpha
k^\alpha  - \mathrm{i}\epsilon_{\mu\nu\alpha\beta}k^\alpha
{k'}^\beta \right) \\
 &  &\!\!\!\!\!\!\!
\!\!\!\!\!\!\!\!\!\!\!\!\!\!\!\!\!\!\!\!
   +8 m s^\alpha \left(k_\nu g_{\alpha\mu}
+ k_\mu g_{\nu\alpha} - g_{\mu\nu} k_\alpha -
\mathrm{i}\epsilon_{\mu\nu\beta\alpha}k^\beta\right)
\end{eqnarray*}\end{footnotesize}
 is a sum of two contributions: one is linear in
the tau lepton momentum  $k'$ the other in its  mass $m$ and its
spin four-vector $s^\mu$.

 The polarization of the $\tau$ measured in
the direction of the four-vector $s_\mu$ is given by
\cite{Bjorken}:
 \begin{footnotesize}
\begin{equation}
\mathcal{P}_{s_\mu}   = \frac{ d\sigma(k,q,s) - d\sigma(k,q,-s)}
{d\sigma(k,q,s) +d\sigma(k,q,-s)} \equiv P_\mu s^\mu
\end{equation}\end{footnotesize}
which defines $P_\mu$ -- the polarization vector of the tau.

We introduce the four-vectors $e_l^\mu$, $e_t^\mu$,
$e_p^\mu$~\cite{Smith,Graczyk} which in the LAB frame have the
following form:
\begin{footnotesize}
$$
e_l^\mu = \frac{1}{m}\left( |\mathbf{k'}|,E_\tau
\frac{\mathbf{k'}}{|\mathbf{k'}|} \right),\quad e_t^\mu=\left(0,
\frac{\mathbf k \times \mathbf k'}{|\mathbf k \times \mathbf
k'|}\right),$$
$$e_p^\mu = \left(0,
\frac{\mathbf e_t \times\mathbf k'}{|\mathbf k'|}\right).
$$
\end{footnotesize}
Writing the polarization four-vector as a linear combination of
$k'$, $e_l$, $e_p$ and $e_t$:
\begin{footnotesize}
\begin{equation}\label{tri} P^\mu
= \alpha {k'}^\mu + e_l^\mu \mathcal{P}_l + e_p^\mu \mathcal{P}_p
+e_t^\mu \mathcal{P}_t
\end{equation}
\end{footnotesize}
defines its longitudinal $\mathcal{P}_l$, perpendicular
$\mathcal{P}_p$ and transverse $\mathcal{P}_t$ components.

To define the degree of polarization it is useful to go into the
rest frame of the tau lepton where the spin four-vector $s_\mu$
has the form:
 {\footnotesize$$ s_\mu = (0,\mathbf{\hat{s}}), \quad
\mathbf{\hat{s}}^2 =1
$$}
Thus the polarization measured in the direction given by
$\mathbf{\hat{s}}$ is equal to: \begin{footnotesize}
$$
\mathcal{P}_\mathbf{s} =-\mathbf{P}\cdot\mathbf{s} = -|\mathbf{P}|
\cos(\beta)
$$\end{footnotesize}
 $\beta$ being the angle between $\mathbf{P}$ and
$\mathbf{\hat{s}}$.  It is easy to notice that in this frame
$\mathbf k'=0$ and (\ref{tri}) implies:
\begin{footnotesize}
\begin{equation} \mathbf P
= \mathbf e_l \mathcal{P}_l + \mathbf e_p \mathcal{P}_p +\mathbf
e_t \mathcal{P}_t
\end{equation}
\end{footnotesize}
The quantity
\begin{footnotesize}
\begin{equation}
\mathcal{P} = |\mathbf{P}| =
\sqrt{\mathcal{P}_l^2+\mathcal{P}_p^2+\mathcal{P}_t^2}.
\end{equation}
\end{footnotesize}
is called the degree of polarization of tau and is frame
independent.

In the LAB frame we choose the coordinate system in which
$\mathcal{P}_t$ vanishes and we obtain the following analytic
formulas for the differential cross section and polarization
vector components \cite{japonczycy}: \begin{footnotesize}
\begin{eqnarray}
\frac{d \sigma}{dE_\tau d\cos(\theta)} & = &  \frac{G^2 p_\tau}{4 \pi M}F \nonumber \\
\label{przekroj} & &
\!\!\!\!\!\!\!\!\!\!\!\!\!\!\!\!\!\!\!\!\!\!\!\!\!\!\!\!
\!\!\!\!\!\!\!\!\!\!\!\!\!\!\!\!\!\!\!\!\!\frac{G^2 p_\tau}{4 \pi
M}\left( 2 F_1 (E_\tau - p_\tau
\cos(\theta))+F_2 \frac{M}{q_0}(E_\tau + p_\tau\cos(\theta)) \right. \nonumber \\
& &\!\!\!\!\!\!\!\!\!\!\!\!\!\!\!\!\!\!\!\!\!\!\!\!\!\!\!\!
\!\!\!\!\!\!\!\!\!\!\!\!\!\!\!\!\!\!\!\!\!\left.+\frac{F_3}{q_0}\left(E
E_\tau + p_\tau^2 -(E+E_\tau)p_\tau \cos(\theta)\right)
-\frac{m^2}{q_0}F_5 \right)
\end{eqnarray}
\begin{eqnarray*}
\mathcal{P}_p & = & \\
& & \!\!\!\!\!\!\!\!\!\!\!\!\!\!\!\!\!\!\!\!\!\!\!\!\!\!\!\!
-\frac{1}{2} \left( 2F_1(p_\tau - E_\tau\cos(\theta))
+ F_2\frac{M}{q_0}(p_\tau+E_\tau \cos(\theta)) \right.\\
& & \!\!\!\!\!\!\!\!\!\!\!\!\!\!\!\!\!\!\!\!\!\!\!\!\!\!\!\!\left.
+\frac{F_3}{q_0}((E+E_\tau)p_\tau - (EE_\tau
+p_\tau^2)\cos(\theta)) - \frac{m^2}{q_0}F_5 \cos(\theta)
\right)/F \\
\mathcal{P}_p& = &\!\! -\frac{m\sin(\theta)}{2}\left ( 2 F_1 -
F_2\frac{M}{q_0} + \frac{E}{q_0}F_3  + \frac{E}{q_0}F_5 \right)/F
\end{eqnarray*}\end{footnotesize}
where $E$ is the neutrino energy,  $E_\tau$ denotes tau  energy,
$q_0=E-E_\tau$ is the energy transfer, $p_\tau$ stands for the
lepton momentum and $M$ for  nucleon mass. It is assumed that
$F_4=0$ and $x F_5 = F_2$.


\begin{figure}
\includegraphics[width=8cm]{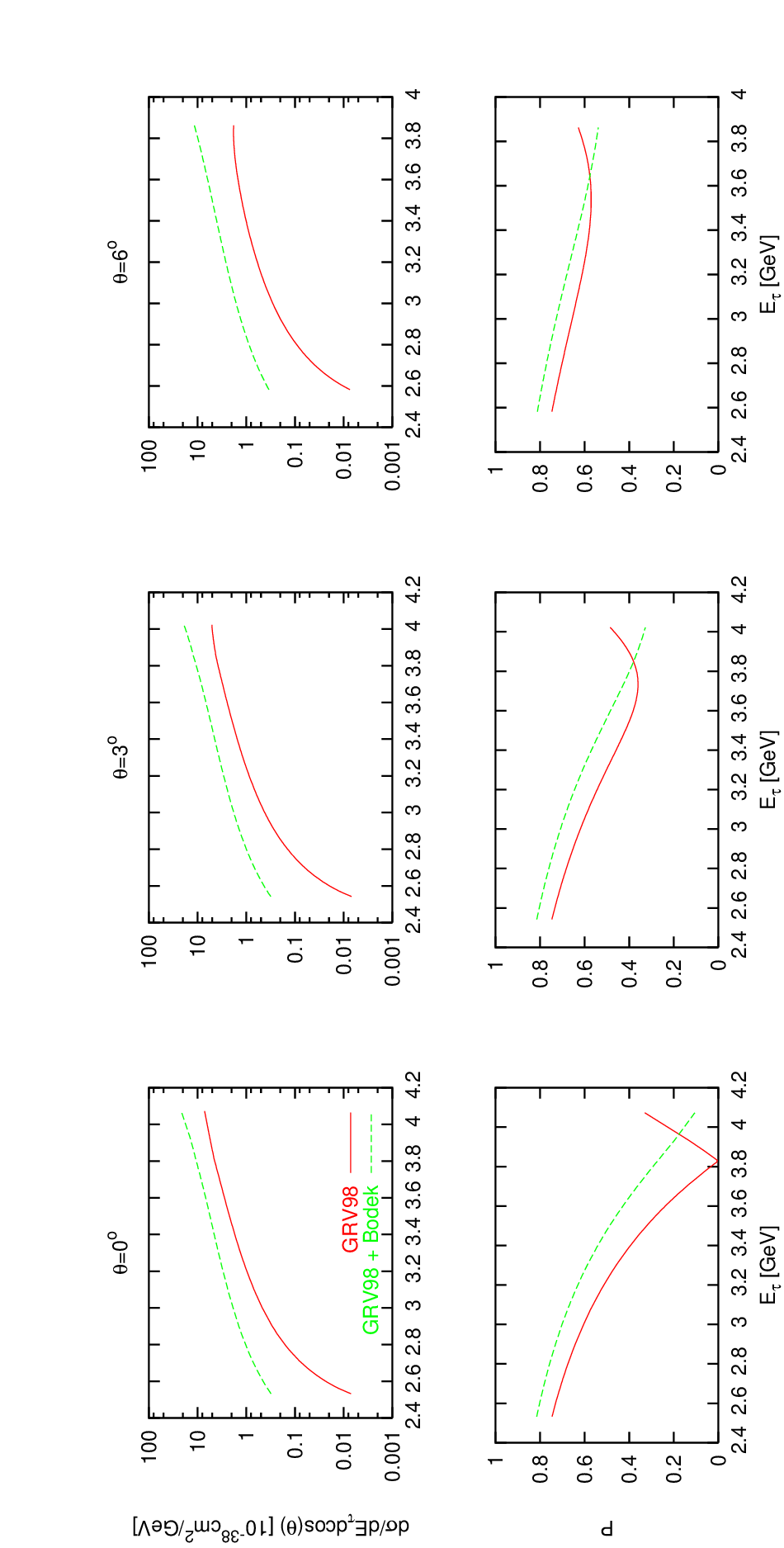}
\scriptsize{ Fig. 1.  In the charts the dependence of the degree
of polarization of $\tau^-$ on its energy and correspond and the
differential cross section  are presented. The calculations are
done for the three scattering angles $\theta = 0^o$, $3^o$, $6^o$.
We compare the plots in the first row where the differential cross
sections are shown with the second row with appropriate plots of
the degree of polarizations. The calculations are done for the
neutrino energy of 4.5~GeV and two sets of PDF's GRV98 -- solid
(red) line and GRV98 with A. Bodek at al. modifications -- dashed
(green) line. \label{rysunek1}}
\end{figure}

\section{Numerical results}
The allowed kinematical region for the DIS formalism is restricted
by the condition for minimal possible hadronic invariant mass $ W
> M+m_\pi $.

The results for the GRV98 (LO) PDF's are obtained with a freeze of
$Q^2$ at 0.8~GeV$^2$. The implementation of the GRV98 with
modifications of A. Bodek is based on the description from the
original paper \cite{Bodek}.

In the figures~1 and~2 we present the dependence of the degree of
the polarization of $\tau^-$ on $E_\tau $ energy. The tau lepton
is produced in neutrino - isoscalar target $(N=\frac{1}{2} (p+n))$
scattering. The calculations are done for 4.5~and~8~GeV neutrino
energy and for three scattering angles $\theta=0^o$, $3^o$, $6^o$.
The charts of the degree of polarization are presented together
with the plots of the corresponding  differential cross sections
given by formula (\ref{przekroj}).

\begin{figure}
\includegraphics[width=8cm]{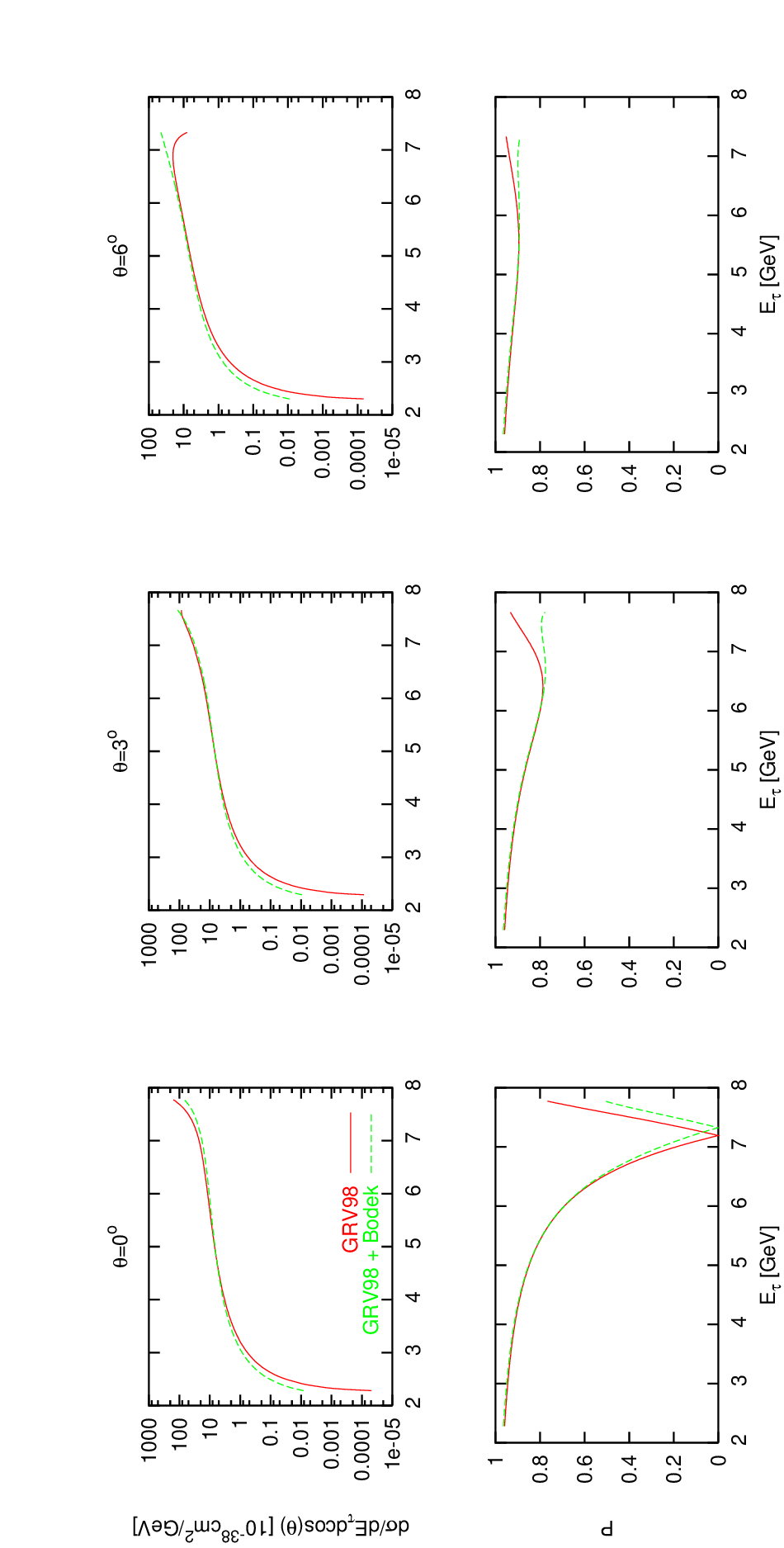}
\scriptsize{Fig. 2. In the charts the dependence of the degree of
polarization of $\tau^-$ on its energy and correspond and the
differential cross section  are presented. The calculations are
done for the three scattering angles $\theta = 0^o$, $3^o$, $6^o$.
We compare the plots in the first row where the differential cross
sections are shown with the second row with appropriate plots of
the degree of polarizations. The calculations are done for the
neutrino energy of 8~GeV and two sets of PDF's GRV98 -- solid
(red) line and GRV98 with A. Bodek at al. modifications -- dashed
(green) line.}\label{rysunek2}
\end{figure}

For the neutrino energy of 4.5~GeV (Fig.~1) the cross sections
which are obtained by using the GRV98 PDF's with modifications are
several times bigger than those calculated by adopting the GRV98
PDF's. However the degree of polarization are comparable.

For the  GRV98 PDF's the degree of the polarization has minimum
which divides the polarization curve into two branches. These
branches correspond two different sings of longitudinal
polarization and at the minimum $\mathcal{P}_l$ vanishes. The
lower (higher) energies of $\tau^-$ correspond to positive
(negative) sign of $\mathcal P_l$ respectively. This effect was
clearly explained by K. Hagiwara at. al.~\cite{japonczycy}. In the
center of mass frame (CM) all tau leptons produced in $\nu N$
reaction are left-handed and are scattered in the all directions.
Performing the Lorentz boost to the LAB frame can transform a
left-handed $\tau^-$ scattered in backward direction in the CM
frame into a right-handed one scattered in the forward directions
in the LAB frame. The degree of polarization of $\tau^-$
calculated by using GRV98 PDF's with modifications have only the
first branch of polarization curves. It means that for the
neutrino energy of 4.5~GeV the leptons are characterized by
positive sign of the helicity.

In the case of the  neutrino energy  of 8~GeV Fig.~2, the
mentioned above effect appear for the both used set of PDF's. The
cross sections as well as the degree of polarization of $\tau^-$
calculated by using the GRV98 PDF's with and without modifications
are  almost the same.

\vskip 0.2cm \textbf{Summary}

The degree of polarization of the $\tau^-$ produced in the $\nu N$
inelastic scattering in the threshold region for the small
scattering angles has the minimum  in the same place where the
differential cross section reaches its maximum.  The $\tau^-$
scattered forward which have energy about close to the neutrino
energy will be only partially polarized.

In the threshold region the PDF's the GRV98 with and without
corrections give very different values of cross section. The
$\mathcal{P}_l$ obtained by using GRV98 with modifications (for
E=4.5) has positive sign. The differences of results given by
application of two investigated sets of PDF's disappear for higher
neutrino energies.

\vskip 0.5cm \textbf{Summary} \vskip 0.1cm

The author would like to thank J. Sobczyk and C. Juszczak for
reading of the manuscript and interesting comments.

\end{document}